\documentclass[a4paper,fleqn]{mnras}
\usepackage{graphicx}
\usepackage[T1]{fontenc}
\usepackage{ae,aecompl}

\title[High-resolution spectroscopy of CEMP-s star CD-50$^\circ$776]{High-resolution 
spectroscopic observations of the new CEMP-s star CD-50$^\circ$776\thanks{Based 
on the observations made with the 2.2m telescope at the European Southern Observatory 
(La Silla, Chile) under the agreement between Observat\'orio Nacional (Brazil) and 
Max-Planck-Institute f\"ur Astronomie}}

\author[Roriz et al.]{M.~Roriz$^{1}$\thanks{E-mail:michelle@on.br}, 
C.~B Pereira$^{1}$\thanks{E-mail:claudio@on.br}, 
N.~A.~Drake$^{1,2}$\thanks{E-mail:drake@on.br}, 
F.~Roig$^{1}$\thanks{E-mail:froig@on.br},
J.~V.~Sales Silva$^{1}$\thanks{E-mail:joaovictor@on.br},
\\ \\
$^{1}$Observat\'orio Nacional/MCTIC, Rua Gen. Jos\'e Cristino, 77, 20921-400, 
Rio de Janeiro, Brazil\\
$^{2}$ Laboratory of Observational Astrophysics, Saint Petersburg State University, 
Universitetski pr. 28, Petrodvoretz 198504, Saint Petersburg,\\ 
Russia\\
}

\date{Accepted xxx. Received xxx; in original form xxxx}
\pubyear{2017}

\begin{document}

\label{firstpage}
\pagerange{\pageref{firstpage}--\pageref{lastpage}}

\maketitle

\begin{abstract} 

\par Carbon enhanced metal poor (CEMP) stars are a particular class of
low metalicity halo stars whose chemical analysis may provide
important contrains to the chemestry evolution of the Galaxy and to
the models of mass transfer and evolution of components in binary
systems. Here, we present a detailed analysis of the CEMP star
CD-50$\degr$776, using high resolution optical spectroscopy. We found
that CD-50$\degr$776 has a metalicity [Fe/H]\,=\,$-2.31$ and a carbon
abundance [C/Fe]\,=\,$+1.21$. Analyzing the s-process elements and the
europium abundances, we show that this star is actually a CEMP-s star,
based on the criteria set in the literature to classify these
chemically peculiar objects. We also show that CD-50$\degr$776 is a
lead star, since it has a ratio [Pb/Ce]\,=\,$+0.97$.  In addition, we
show that CD-50$\degr$776 develops radial velocity variations that may
be attributed to the orbital motion in a binary system.  The abundance
pattern of CD-50$\degr$776 is discussed and compared to other CEMP-s
stars already reported in the literature to show that this star is a
quite exceptional object among the CEMP stars, particularly due to its
low nitrogen abundance. Explaining this pattern may require to improve
the nucleosynthesis models, and the evolutionary models of mass
transfer and binary interaction.

\end{abstract}


\section{Introduction}

\par During the last decades, considerable theoretical and
observational efforts have been made to investigate the formation and
evolution of the chemistry of the Galaxy through the study of its halo
stars. The chemical composition of the halo stars is important because
it can provide information not only about the early stages of the
Galaxy formation, but also about some sites where nucleosynthesis of
several elements took place, thus providing significant evidence to
describe the initial stages of galactic nucleosynthesis. In order to
carry on such studies, an adequate sample of halo stars must be
selected. After the first surveys of low metallicity stars initiated
by Bond et al. (1970, 1980), Bidelman (1981) and Bidelman \&
MacConnell (1973), where the metallicity limit was around $-2.6$
(Frebel \& Norris, 2013, Beers et al. 2014), the surveys by Beers et
al. (1985) and Christlieb et al. (2001) significantly increased the
known number of low metalicity stars, including several candidates
with metallicities less than $-2.0$.

\par Following these surveys, spectroscopic studies revealed that some
of the metal-poor stars from Beers et al. (1985) were also carbon-rich
objects (Beers et al. 1992). Before 2005, high-resolution
spectroscopic analysis of these stars confirmed the carbon-rich nature
for some of them. In addition, it was noted that some of these stars
were also enriched in either the r-, s- or r/s-procesess (McWilliam et
al. 1995; Sneden et al. 1994, 1996, 2003a, 2003b; Barbuy et al. 1997,
Norris et al. 1997a, 1997b, Bonifacio et al. 1998, Hill et al. 2000,
2002, Sivarani et al. 2004).  In 2005, Rossi et al. (2005) using
medium-resolution spectra of the stars in the samples of Beers et
al. (1985) and Christlieb et al. (2001), noted the high frequency
occurrence of the carbon-rich stars among the metal-poor stars, also
known as CEMP (carbon enhanced metal poor) stars.  According to
Lucatello et al. (2006), 20\% of the stars with metallicities down to
$-2.0$ are CEMP stars. CEMP stars have been found in all the
metallicity range from $-2.0$ to $-4.0$, with increasing frequency
towards the lower metallicities (Lucatello et al. 2006). In view of
this, several astrophysical sites for the origin of the carbon
overabundances have been proposed (see Beers \& Christlieb 2005 for a
discussion).

\par Beers \& Christlieb (2005) (but see also Masseron et al. 2010)
proposed that CEMP stars could be distinguished according to their
barium and europium abundances, and also according to their [Ba/Eu]
ratio. After their study, the CEMP stars were divided in CEMP-s,
CEMP-r, CEMP-r/s and CEMP-no stars according to the heavy elements
abundance pattern. The majority of CEMP stars are CEMP-s stars (Aoki
et al. 2007). The most likely explanation for the observed excess of
carbon and s-process elements in CEMP-s stars is the mass-transfer,
just like in the CH stars and the barium stars. This conclusion is
supported by the radial-velocity variations observed in several CEMP-s
stars (Hansen et al. 2016). Therefore, detailed abundance analysis of
CEMP-s stars is important to set observational constraints to the
physics of mass-transfer (in the case of CEMP-s binaries), and also to
the nucleosynthesis models.

\par In this work we present the spectroscopic analysis of a new
CEMP-s star: CD-50$^\circ$776. CD-50$^\circ$776 came to our attention
during our high-resolution spectroscopy survey started in 1999, during
the first agreement between Observat\'orio Nacional and the European
Southern Observatory, with the aim to search for halo chemically
peculiar stars. Later on, we also searched for metal-poor
hypervelocity candidate stars, following our analysis of
CD-62$^\circ$1346, a CH hypervelocity star candidate (Pereira et
al. 2012), and the analysis of two metal-poor red-horizontal-branch
stars: CD-41$^\circ$115048 and HD 214362 (Pereira et al. 2013). To
select these peculiar stars, we search over several surveys from the
literature. In particular, CD-50$^\circ$776 was selected from the work
of Bidelman \& MacConnel (1973), whose stars sample was later
investigated by Norris et al. (1985) and Beers et al. (2014, 2017).
In particular, Beers et al. (2014), based on a medium-resolution
spectrum, determined the metallicity and the [C/Fe] ratio of this
star, obtaining values of $-2.23$ and $+1.81$, respectively.  Here, we
show that CD-50$^\circ$776 is in fact a CEMP star with an excess of
the elements created by the s-process, without europium
enrichment. Therefore it can be classified as a CEMP-s star.
CD-50$^\circ$776 is the second brightest CEMP-s star known to date,
with $V=10.05$ (the brightest one is HD 196944 with $V=8.4$).  The
present work is based on the analysis of high-resolution spectra of
CD-50$^\circ$776 to determine its metallicity and abundance pattern.

\section{Observations} 

\par The high-resolution spectra analyzed in this work were obtained
with the Feros (Fiberfed Extended Range Optical Spectrograph)
spectrograph (Kaufer et al. 1999), that was initially coupled to the
1.52\,m telescope and later to the 2.2\,m telescope of ESO, at La
Silla (Chile). Two observations were done for CD-50$^\circ$776.  One,
on October 26, 1999 and another one on September 25, 2016.  The
exposures were 3600 and 2400 secs, respectively.  Feros consists of a
CCD detector of $2048\times 4096$ pixels having each pixel a size of
$15\,\mu m$. Feros has spectral coverage between 3\,900\AA\, and
9\,200\AA\, distributed over 39 orders with a resolution of
48\,000. The spectral reduction was made following a standard
procedure, which includes bias subtraction, flat-fielding, spectral
order extraction and wavelength calibration.  All this procedure has
been done using the MIDAS reduction pipeline.

\section{Analysis \& results}

\par The atomic absorption lines used for the determination of
atmospheric parameters are basically the same as which were used in
the study of other chemically peculiar stars (Pereira \& Drake
2009). Table 1 shows the Fe\,{\sc i} and Fe\,{\sc ii} lines used to
determine these parameters. The $\log gf$ values for the Fe\,{\sc i}
and Fe\,{\sc ii} lines were taken from Lambert et al. (1996).

\subsection{Determination of the atmospheric parameters}  

\par Our analysis was done using the spectral analysis code MOOG
(Sneden 1973) and the model atmospheres of Kurucz (1993). The latest
version of MOOG includes routines for the calculation of the
Rayleigh-scattering contribution to the continuous opacity, as
described in Sobeck et al. (2011). The temperature was obtained after
searching for a zero slope of the relation between the iron abundances
based on Fe\,{\sc i} lines and the excitation potential while the
microturbulent velocity was obtained after searching for a zero slope
of the relation between the iron abundances based on the same Fe\,{\sc
  i} lines and the reduced equivalent width
($W_{\lambda}/{\lambda}$). This procedure also provides the
metallicity of the star.  The surface gravity of the star was obtained
by means of the ionization equilibrium, which means that we should 
find a solution until the abundance of Fe\,{\sc i} and Fe\,{\sc ii}
become equal.

\par The final atmospheric parameters derived for CD-50$^\circ$776 are
given in Table~2. Table 2 also shows the values derived from
previous spectroscopic observations of CD-50$^\circ$776 conducted by
Ryan \& Deliyannis (1998) and Beers et al. (2014). The three
atmospheric parameters given in Beers et al. (2014), labelled as 2a,
2b and 2c, differ according to the techniques used by these authors to
obtain them.  We note that our atmospheric parameters are in a good
agreement either in temperature or surface gravity, depending on the
specific technique used by Beers et al. (2014). A model with the
highest temperature implies a change of $+0.5$ dex in the carbon
abundance compared to our results. The model with $\log g=3.0$ does
not allow a good fit in the region of the C$_{2}$ molecule, at
5165\AA.

\par The errors reported in our effective temperature ($T_{\rm eff}$)
and microturbulent velocity ($\xi$) were set from the uncertainty in
the slope of the Fe\,{\sc i} abundance {\sl versus.} excitation
potential and {\sl versus.} $W_{\lambda}/{\lambda}$, respectively.  For the gravity,
the error was estimated until the mean abundances of Fe\,{\sc i} and Fe\,{\sc ii} 
differ by 1$\sigma$ of the standard deviation of the [Fe\,{\sc i}/H] mean value.

\begin{table*}
\caption{Observed Fe\,{\sc i} and Fe\,{\sc ii} lines.}
\begin{tabular}{ccccc}\hline
Element &  $\lambda$\,(\AA) &  $\chi$\,(eV) & $\log gf$ & $W_\lambda$\,(m\AA) \\ 
\hline
Fe\,{\sc i}  &  4187.050  &     2.450  &   $-$0.55  &   88\\
 &  4233.610  &     2.480  &   $-$0.60  &   77\\
 &  4494.570  &     2.200  &   $-$1.14  &   74\\
 &  4531.160  &     1.490  &   $-$2.15  &   73\\
 &  4871.330  &     2.860  &   $-$0.36  &   85\\
 &  5110.413  &     0.000  &   $-$3.76  &   82\\
 &  5171.596  &     1.485  &   $-$1.76  &   87\\
 &  5194.942  &     1.557  &   $-$2.09  &   73\\
 &  5198.711  &     2.223  &   $-$2.14  &   34\\
 &  5202.336  &     2.176  &   $-$1.84  &   57\\
 &  5242.491  &     3.634  &   $-$0.97  &   19\\
 &  5250.209  &     0.121  &   $-$4.92  &   19\\
 &  5281.790  &     3.038  &   $-$0.83  &   49\\
 &  5302.307  &     3.283  &   $-$0.74  &   43\\
 &  5307.361  &     1.608  &   $-$2.97  &   27\\
 &  5339.929  &     3.266  &   $-$0.68  &   48\\
 &  5341.024  &     1.608  &   $-$1.95  &   79\\
 &  5364.871  &     4.445  &      0.23  &   30\\
 &  5367.467  &     4.415  &      0.44  &   37\\
 &  5369.962  &     4.371  &      0.54  &   42\\
 &  5389.479  &     4.415  &   $-$0.25  &   11\\
 &  5393.168  &     3.241  &   $-$0.72  &   44\\
 &  5400.502  &     4.371  &   $-$0.10  &   21\\
 &  5410.910  &     4.473  &      0.40  &   29\\
 &  5445.042  &     4.386  &      0.04  &   28\\
 &  5497.516  &     1.011  &   $-$2.84  &   72\\
 &  5569.618  &     3.417  &   $-$0.49  &   52\\
 &  5572.842  &     3.396  &   $-$0.28  &   60\\
 &  5576.089  &     3.430  &   $-$0.85  &   36\\
 &  5638.262  &     4.220  &   $-$0.72  &   12\\
 &  6065.482  &     2.608  &   $-$1.53  &   46\\
 &  6136.615  &     2.453  &   $-$1.40  &   61\\
 &  6137.692  &     2.588  &   $-$1.40  &   54\\
 &  6191.558  &     2.433  &   $-$1.40  &   57\\
 &  6200.313  &     2.605  &   $-$2.44  &   11\\
 &  6230.723  &     2.559  &   $-$1.28  &   63\\
 &  6252.555  &     2.403  &   $-$1.72  &   57\\
 &  6265.130  &     2.180  &   $-$2.55  &   23\\
 &  6322.686  &     2.588  &   $-$2.43  &   14\\
 &  6393.601  &     2.433  &   $-$1.43  &   57\\
 &  6411.649  &     3.653  &   $-$0.66  &   32\\
 &  6421.351  &     2.279  &   $-$2.01  &   42\\
 &  6430.846  &     2.176  &   $-$2.01  &   49\\
 &  6592.914  &     2.723  &   $-$1.47  &   40\\
 &  6593.871  &     2.437  &   $-$2.42  &   19\\\hline
Fe\,{\sc ii} &  4515.339  &     2.840  &   $-$2.45  &   37\\
 &  4583.837  &     2.810  &   $-$1.80  &   71\\
 &  5197.559  &     3.230  &   $-$2.25  &   35\\
 &  5234.619  &     3.221  &   $-$2.24  &   41\\
 &  5284.098  &     2.891  &   $-$3.01  &   18\\
 &  5425.247  &     3.199  &   $-$3.21  &   10\\
 &  5534.834  &     3.245  &   $-$2.77  &   16\\
 &  6247.545  &     3.891  &   $-$2.34  &   11\\
 &  6432.682  &     2.891  &   $-$3.58  &    7\\\hline
\end{tabular}
\end{table*}

\begin{table} 
\caption{Atmospheric parameters of CD-50$^\circ$776.}
\begin{tabular}{lcc}\hline
Parameter & Value &  Ref. \\\hline
$T_{\rm eff}$ (K)        & 4\,900$\pm$ 60   & 1  \\
                       & 5\,305           & 2a \\
                       & 5\,000           & 2b \\
                       & 5\,176           & 2c \\
                       & 5\,000           & 3  \\\hline
                       
$\log g$ (dex)         & 2.1$\pm$0.2      & 1  \\
                       & 2.22             & 2a \\
                       & 3.0              & 2b \\
                       & 2.22             & 2c \\\hline

[Fe/H] (dex)           & $-$2.31$\pm$ 0.08  & 1   \\
                       & $-$2.39            & 2a \\
                       & $-$2.23            & 2b \\
                       & $-$2.52            & 2c \\
                       & $-$2.23            & 3 \\\hline

$\xi$ (km\,s$^{-1}$)    & 1.5 $\pm$ 0.3      & 1 \\\hline
\end{tabular}
\par References for Table 2.
\par 1: This work;
\par 2: Beers et al. (2014);
\par 3: Ryan \& Deliyannis (1998);
\end{table}

\subsection{Abundance analysis}                                        

\par The abundance pattern of CD-50$^\circ$776 was determined using
either equivalent width measurements of selected atomic lines and
using the spectral synthesis technique. We used the solar abundances
of Grevesse \& Sauval (1998) as a reference. For iron it was used the
solar abundance of $\log \varepsilon$(Fe)\,=\,7.52.  Table~3 shows the
atomic lines used to derive the abundances of the elements, with their
respective equivalent width measurements.  The derived abundances are
given in Table~4. For the elements whose abundances were derived using
spectral synthesis technique they are labelled as {\sl syn}.

\par The abundances of the light elements, carbon and nitrogen, were
determined by applying a spectrum synthesis technique in the local
thermodynamic equilibrium (LTE). For carbon, we used the CH lines of
the $A^2\Delta - X^2B$ system at $\sim 4365$\,\AA, the C$_{2}$ (0,0)
band head of the Swan system $d^3\Pi_{g} - a^3\Pi_{u}$ at 5165\,\AA,
and the C$_{2}$ (0,1) band head of the Swan system $d^3\Pi_{g} -
a^3\Pi_{u}$ at 5635\,\AA.

\par For nitrogen we used the $B^2\Sigma - X^2\Sigma$ violet system
band head at 3883\,\AA\, with line list provided by VALD. The (2,\,0)
band of the CN red system $A^2\Pi - X^2\Sigma$ in the
7994--8020\,\AA\, often used by us to determine the nitrogen
abundance, is not visible in this star. We did not detect the oxygen
forbidden line at 6300.0\,\AA\, Therefore we assume that
[O/Fe]\,=\,$+0.50$, which is a typical value for a star of this
metallicity (Masseron et al. 2006).  We also check our derived
nitrogen abundance using a different linelist for the CN band at
3883\,\AA\, given by Jonsell et al. (2006) and Sneden et al. (2014),
and the results were basically the same as using the linelist given by
VALD.

\par The abundances of barium, europium, cobalt, lead and praseodymium
were also determined by means of spectral synthesis technique.  The
determination of barium abundance was obtained using the Ba\,{\sc ii}
lines at $\lambda 4554.0$, $\lambda 4934.1$, $\lambda 5853,7$, and
$\lambda 6141.7$\,\AA. Hyperfine and isotope splitting were taken from
McWilliam (1998).  The europium abundance was found using the line of
Eu\,{\sc ii} at $\lambda 4129.75$\,\AA\, and the hyperfine splitting
from Mucciarelli et al. (2008).  The cobalt abundance was derived
using the Co\,{\sc i} line at $\lambda 4121.33$\,\AA, where the
hyperfine splitting was taken from McWilliam et al. (1995).  The lead
abundance was derived from the Pb\,{\sc i} line at $\lambda
4057.81$\,\AA.  The line data, which include isotopic shifts and
hyperfine splitting, were taken from van Eck et al. (2003).  The
abundance of praseodymium was obtained through spectral synthesis
technique using the lines at 5259.73\AA\, and 5322.77\AA. The
hyperfine splitting was taken from Sneden et al. (2009).

\par Figures~1, 2, 3, 4 and 5 show the observed and synthetic spectra
for the spectral regions where the abundances of carbon, the
$^{12}$C/$^{13}$C isotopic ratio, nitrogen, lead and europium were
obtained.

\begin{figure} 
\includegraphics[width=\columnwidth]{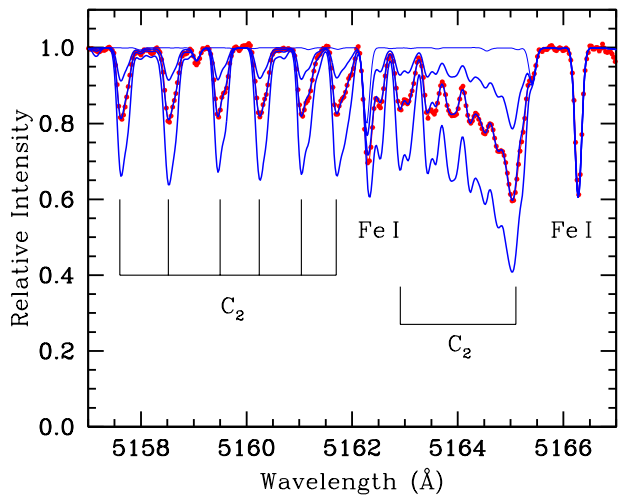}
\caption{Observed (dotted red points) and synthetic (solid blue lines)
spectra in the region containing the C$_{2}$ molecular lines at 
$\lambda 5165$\,\AA. From top to bottom, we show the syntheses 
for the carbon  abundances of $\log\varepsilon$(C)\,=\,7.22, 7.42 (adopted), and 7.62. 
Other absorption lines are also indicated. The light blue line shows the 
synthetic spectrum calculated without the C$_{2}$ molecular lines contribution.}
\end{figure}

\begin{figure} 
\includegraphics[width=\columnwidth]{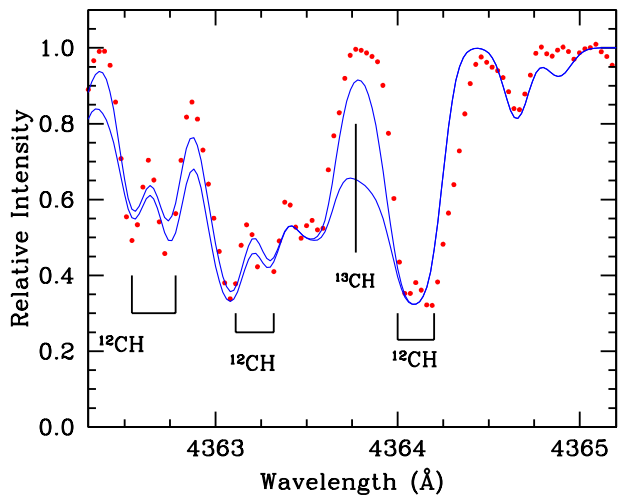}
\caption{Observed (dotted red points) and synthetic (solid blue lines)
spectra in the region containing the CH molecular lines 
around 4364\,\AA. In the synthetic spectra (from top to bottom) we show the syntheses 
for the $^{12}$C/$^{13}$C\,=\,64 and 8.}
\end{figure}

\begin{figure} 
\includegraphics[width=\columnwidth]{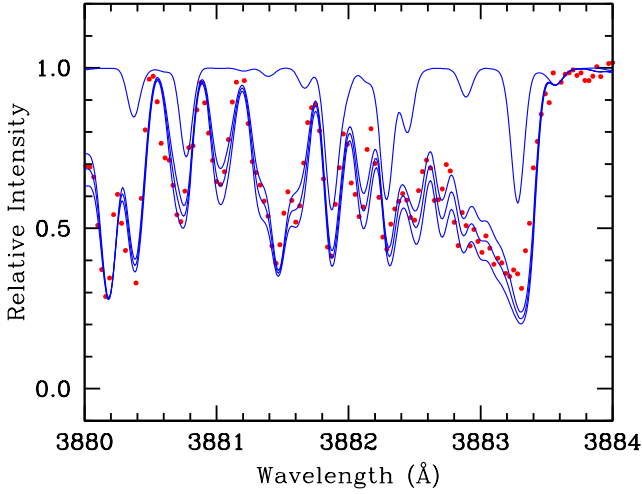}
\caption{Observed (dotted red points) and synthetic (solid blue lines)
spectra between 3880\,\AA\, and 3884\,\AA.\, From top to bottom, we show 
the syntheses for the nitrogen abundances of $\log\varepsilon$(N)\,=\,5.22, 5.52 (adopted)
and 5.82. The upper line shows the spectrum without contribution of the CN lines.}
\end{figure}

\begin{figure} 
\includegraphics[width=\columnwidth]{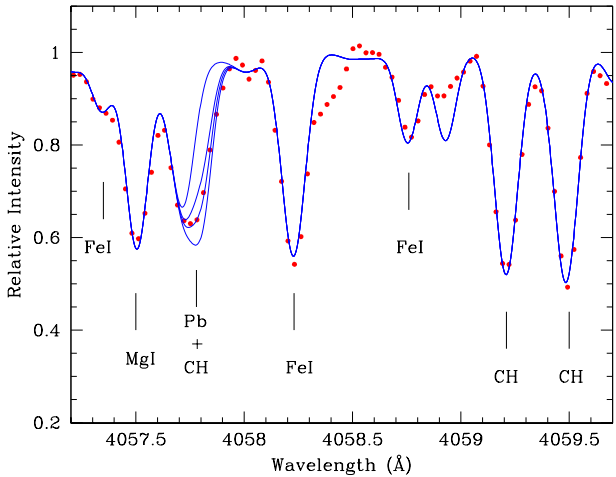}
\caption{Observed (dotted red points) and synthetic (solid blue lines)
spectra in the region of the Pb\,{\sc i} at 4057.8\AA\,.
From top to bottom, we show the syntheses without contribution of the lead 
and the lead abundances of $\log\varepsilon$(Pb)\,=\,1.33, 1.53 (adopted), and 
1.73. Other absorption lines are indicated.}
\end{figure}

\begin{figure} 
\includegraphics[width=\columnwidth]{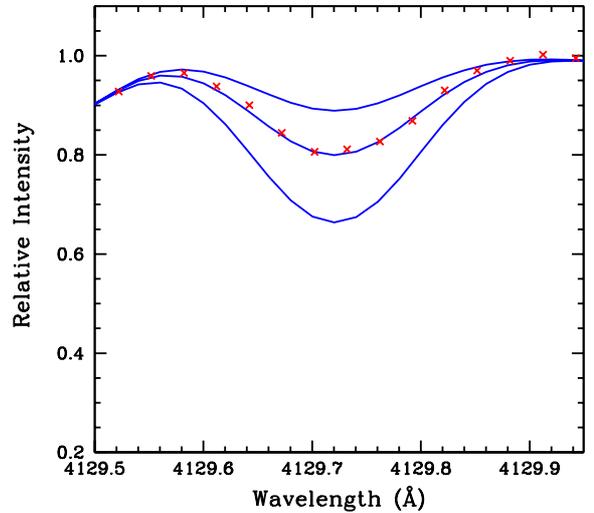}
\caption{Observed (dotted red points) and synthetic (solid blue lines)
spectra for the Eu\,{\sc ii} line at 4129.7\,\AA\,. From top to bottom
we show the syntheses for the europium abundances of
$\log\varepsilon$(Eu)\,=\,$-$1.79, $-$1.49 (adopted) and
$-$1.19.} 
\end{figure}

\begin{table*}
\caption{Other lines studied}
\begin{tabular}{cccccc}\hline
$\lambda$\,(\AA) & Species & $\chi$\,(eV) & $\log gf$ & Ref. & $W_{\lambda}$\,(m\AA) \\\hline
   5688.22  & Na\,{\sc i}  &   2.10  &  $-$0.40  & PS  &   10\\\hline

   4057.51  & Mg\,{\sc i} & 4.35     & $-$0.89 &  N96 & 56\\
   4571.10  &  & 0.00   &   $-$5.61  &  N96 &    62\\
   4702.99  &  & 4.35   &   $-$0.38  &  N96 &    94\\
   5528.42  &  & 4.34   &   $-$0.36  &  R99 &    86\\
   5711.10  &  & 4.34   &   $-$1.75  &  R99 &    22\\\hline

   5581.80   & Ca\,{\sc i}&  2.52    &   $-$0.67 &  C2003 &  26\\
   5601.29   & &  2.52   &   $-$0.52 &  C2003 &    38\\
   5857.46   & &  2.93   &      0.11 &  C2003 &    40\\
   6102.73   & &  1.88   &   $-$0.79 &  D2002 &    51\\
   6122.23   & &  1.89   &   $-$0.32 &  D2002 &    78\\
   6162.18   & &  1.90   &   $-$0.09 &  D2002 &    90\\
   6166.44   & &  2.52   &   $-$1.14 &  R03   &    10\\
   6169.04   & &  2.52   &   $-$0.80 &  R03   &    21\\
   6169.56   & &  2.53   &   $-$0.48 &  DS91  &    30\\
   6439.08   & &  2.52   &      0.47 &  D2002 &    76\\
   6493.79   & &  2.52   &   $-$0.11 &  DS91  &    49\\
   6499.65   & &  2.52   &   $-$0.81 &  C2003 &    13\\
   6717.69   & &  2.71   &   $-$0.52 &  C2003 &    23\\\hline
   4512.74   & Ti\,{\sc i} & 0.84  &  $-$0.48 &  MFK &  26\\
   4518.03   & &  0.83   &  $-$0.32 &  MFK    &  30\\
   4533.25   & &  0.85   &  $+$0.48 &  MFK    &  60\\
   4548.77   & &  0.83   &  $-$0.35 &  MFK    &  23\\
   4555.49   & &  0.85   &  $-$0.49 &  MFK    &  16\\
   4981.72   & &  0.84   &     0.50 &  MFK    &  61\\
   4999.51   & &  0.83   &     0.25 &  MFK    &  58\\
   5016.17   & &  0.85   &  $-$0.57 &  MFK    &  16\\
   5022.87   & &  0.83   &  $-$0.43 &  MFK    &  22\\
   5173.75   & &  0.00   &  $-$1.12 &  MFK    &  30\\
   5210.39   & &  0.05   &  $-$0.88 &  MFK    &  43\\\hline
   4254.35   & Cr\,{\sc i} &    0.00  &  $-$0.09 &  S2007 & 112\\
   4496.84   & &  0.94   &   $-$1.14  & S2007  &    25\\
   5206.04   & &  0.94   &      0.02  & S2007  &    84\\
   5247.57   & &  0.96   &   $-$1.60  & S2007  &    15\\
   5296.70   & &  0.98   &   $-$1.37  & S2007  &    25\\
   5298.28   & &  0.98   &   $-$1.14  & S2007  &    29\\
   5345.81   & &  1.00   &   $-$0.95  & S2007  &    41\\
   5348.33   & &  1.00   &   $-$1.22  & S2007  &    28\\
   5409.80   & &  1.03   &   $-$0.67  & S2007  &    51\\\hline

   4904.41 & Ni\,{\sc i} &   3.54 &  $-$0.24 & W2014 & 19\\
   4648.65 & &    3.42  &  $-$0.09  &  W2014 &  20\\
   4756.52 & &    3.48  &  $-$0.27  &  W2014 &  17\\
   5035.36 & &    3.64  &     0.29  &  W2014 &  25\\
   5476.90 & &    1.83  &  $-$0.78  &  W2014 &  74\\
   5892.88 & &    1.99  &  $-$1.92  &  W2014 &  15\\
   6108.11 & &    1.68  &  $-$2.69  &  W2014 &  10\\
   6482.80 & &    1.94  &  $-$2.63  &  MFK   &  11\\
   6643.64 & &    1.68  &  $-$2.03  &  W2014 &  19\\
   6767.77 & &    1.83  &  $-$2.17  &  W2014 &  18\\
   7788.93 & &    1.95  &  $-$1.99  &  W2014 &  13\\\hline

   4810.53   & Zn\,{\sc i} &  4.06   &  $-$0.17 & BG80 &    26\\\hline

   4215.52   & Sr\,{\sc ii}&  0.00   &  $-$0.17 & N96  &   142\\\hline

   4883.68   & Y\,{\sc ii}&  1.08  &    0.07 & H82 &  44\\
   5087.43   & &  1.08  &    $-$0.17 &  H82 &   31\\
   5200.41   & &  0.99  &    $-$0.57 &  H82 &   16\\
   5205.72   & &  1.03  &    $-$0.34 &  S96 &   21\\\hline
\end{tabular}
\end{table*}

\setcounter{table}{2}
\begin{table*}
\caption{continued}
\begin{tabular}{cccccc}\hline

$\lambda$\,(\AA) & Species & $\chi$\,(eV) & $\log gf$ & Ref. & $W_{\lambda}$\,(m\AA) \\\hline
   4050.32   & Zr\,{\sc ii} &  0.71  &  $-$1.06  & L2006  &  19\\
   4208.98   &  &  0.71  &  $-$0.51  & L2006  &  51\\
   4554.80   &  &  0.80  &  $-$1.18  & L2006  &  12\\\hline

   4086.71   & La\,{\sc ii} & 0.00   &  $-$0.16 & L01 &  49\\
   6390.48   & &  0.32   &   $-$1.41 & L01 &    7\\\hline
 
   4486.91   & Ce\,{\sc ii}&  0.29   &   $-$0.18 & L09 &  29\\
   4539.74   & &  0.33   &   $-$0.08 & L09 &    27\\
   4562.37   & &  0.48   &      0.21 & L09 &    37\\
   4628.16   & &  0.52   &      0.14 & L09 &    33\\
   5187.46   & &  1.21   &      0.17 & L09 &     6\\
   5274.24   & &  1.04   &      0.13 & L09 &    12\\
   5330.58   & &  0.87   &   $-$0.40 & L09 &     7\\\hline

   4820.34   & Nd\,{\sc ii}& 0.20  &  $-$0.92 & DH & 17\\
   4825.48   & &  0.18   &   $-$0.42   & DH &    24\\
   4914.38   & &  0.38   &   $-$0.70   & DH &    14\\
   5192.61   & &  1.14   &      0.27   & DH &    12\\
   5212.36   & &  0.20   &   $-$0.96   & DH &    12\\
   5234.19   & &  0.55   &   $-$0.51   & DH &    10\\
   5249.58   & &  0.98   &      0.20   & DH &    16\\
   5255.51   & &  0.20   &   $-$0.67   & DH &    16\\
   5293.16   & &  0.82   &      0.10   & DH &    21\\
   5319.81   & &  0.55   &   $-$0.14   & DH &    21\\\hline

  4318.94    & Sm\,{\sc ii} & 0.28   & $-$0.25 & L06 & 14 \\  
  4424.32    & &  0.48  &      0.14  &  L06 &  23 \\
  4467.34    & &  0.66  &     0.15   &  L06 &  10 \\
  4566.20    & &  0.33  &   $-$0.59  &  L06 &   7 \\
  4815.80    & &  0.18  &   $-$0.82  &  L06 &   6 \\\hline
\end{tabular}
\par References for Table 3.
\par BG80: Biemont \& Godefroid (1980); C2003: Chen et al. (2003);
\par D2002: Depagne et al. (2002); DS91: Drake \& Smith (1991);
\par DH: Den Hartog et al. (2003); E93: Edvardsson et al. (1993)
\par GS: Gratton \&  Sneden (1988); H82: Hannaford et al. (1982); 
\par L01: Lawler et al. (2001); L2006: Ljung et al. (2006);
\par L06: Lawler et al. (2006); L09: Lawler et al. (2009); 
\par McW: McWilliam et al. (1995); MFK : Martin et al. (2002); 
\par N96: Norris et al (1996); PS: Preston \& Sneden (2001); 
\par R99: Reddy et al. (1999); R03: Reddy et al. (2003); 
\par S2007: Sobeck et al. (2007); S96: Smith et al. (1996); 
\par W2014 : Wood et al. (2014); Z2009 : Zhang et al. (2009);
\end{table*}

\begin{table} 
\caption{Chemical abundances derived for CD-50$^\circ$776 in the scale 
$\log\varepsilon$(H)\,=\,12.0, and in the notations [X/H] and [X/Fe].} 
\begin{tabular}{lcccc}\hline
Species & n & $\log\varepsilon$ & [X/H] & [X/Fe]  \\ \hline 
C\,(C$_{2}$)  & syn(3)  & 7.42$\pm$0.10    & $-$1.10 & $+$1.21 \\
N\,(CN)      & syn(1)  & 5.52             & $-$2.40 & $-$0.09 \\
Na\,{\sc i}  &  1      & 4.09             & $-$2.04 & $+$0.07 \\
Mg\,{\sc i}  &  5      & 5.74$\pm$0.15    & $-$1.84 & $+$0.47 \\
Ca\,{\sc i}  & 13      & 4.54$\pm$0.08    & $-$1.82 & $+$0.49 \\
Ti\,{\sc i}  & 11      & 2.91$\pm$0.09    & $-$2.11 & $+$0.20 \\
Cr\,{\sc i}  &  9      & 3.17$\pm$0.06    & $-$2.50 & $-$0.19 \\
Co\,{\sc i}  & syn(1)  & 2.52             & $-$2.40 & $-$0.09 \\
Ni\,{\sc i}  & 11      & 3.99$\pm$0.14    & $-$2.26 & $+$0.05 \\
Zn\,{\sc i}  &  1      & 2.52             & $-$2.08 & $+$0.23 \\
Sr\,{\sc ii}  &  1     & 0.74             & $-$2.23 & $+$0.08 \\
Y\,{\sc ii}   &  4     & 0.04$\pm$0.10    & $-$2.20 & $+$0.11 \\
Zr\,{\sc ii}  &  3     & 0.76$\pm$0.08    & $-$1.84 & $+$0.47 \\
Ba\,{\sc ii}  & syn(3) & 0.83$\pm$0.10    & $-$1.30 & $+$1.01 \\
La\,{\sc ii}  &  2     & $-$0.28          & $-$1.45 & $+$0.86 \\
Ce\,{\sc ii}  &  7     & 0.19$\pm$0.08    & $-$1.39 & $+$0.92 \\
Pr\,{\sc ii}  & syn(2) & $-$0.82          & $-$1.53 & $+$0.78 \\ 
Nd\,{\sc ii}  & 10     & 0.04$\pm$0.11    & $-$1.46 & $+$0.85 \\
Sm\,{\sc ii}  &  5     & $-$0.61$\pm$0.10 & $-$1.62 & $+$0.69 \\
Eu\,{\sc ii}  & syn(1) & $-$1.49          & $-$2.00 & $+$0.31 \\
Pb\,{\sc i}   & syn(1) & 1.53             & $-$0.42 & $+$1.89 \\\hline
\end{tabular}
\par $^{12}$C/$^{13}$C\,$\geq$\,64
\end{table}

\subsection{Abundance uncertainties}    

\par The uncertainties in the abundances of CD-50$^\circ$776 are given
in Table~5.  The uncertainties due to the errors of $T_{\rm eff}$,
$\log g$, $\xi$, and metallicity were estimated by changing these
parameters one at a time by their standard errors given in Table~2.
The final uncertainties of the abundances were calculated as the root
squared sum of the individual uncertainties due to the errors in each
atmospheric parameter and also in the equivalent widths under the
assumption that these individual uncertainties are independent.

\par For the elements analyzed via spectrum synthesis we used the same
technique, varying the atmospheric parameters and then computing
independently the abundance changes introduced by them.  Uncertainties
in the carbon abundances also result in variations of the nitrogen
abundances, since the CN molecular lines were used for the nitrogen
abundance determination. For carbon and nitrogen, typical
uncertainties are 0.10 and 0.20, respectively.  In Table~5, we see
that the neutral elements are more sensitive to the temperature
variations, while singly-ionized elements are more sensitive to the
$\log g$ variations.  For the elements whose abundance is based on
stronger lines, such as strontium, the error introduced by the
microturbulence is important.  Finally, we observe that the abundances
of carbon and nitrogen are weakly sensitive to the variations of the
microturbulent velocity.

\begin{table*}
\caption{Abundance uncertainties for CD-50$^\circ$776. Columns 2 to 6 give the 
variation of the abundances caused by the variation in $T_{\rm eff}$, $\log g$, $\xi$, 
[Fe/H], and equivalent widths measurements ($W_\lambda$), respectively. 
The 7th column gives the compounded r.m.s. 
uncertainty from the 2nd to 6th columns. The last column gives the 
abundances dispersion observed among the lines for those elements with more than 
three lines available.}
\begin{tabular}{lccccccc}\hline
Species & $\Delta T_{\rm eff}$ & $\Delta\log g$ & $\Delta\xi$ & $\Delta$[Fe/H] & 
$\Delta W_{\lambda}$ & $\left( \sum \sigma^2 \right)^{1/2}$ & $\sigma_{\rm obs}$\\
$_{\rule{0pt}{8pt}}$ & $+60$\,K & $+0.2$ & $+$0.3 km\,s$^{-1}$ & $+$0.1 & $+$3 m\AA &  \\\hline
C              &  $+$0.12 &  $-$0.03 & 0.00    &  0.00   &  ---     &  0.12  &  --- \\
N              &  $+$0.16 &  $-$0.10 & 0.00    &  0.00   &  ---     &  0.19  &  --- \\
Fe\,{\sc i}    & $+$0.08  &    0.00  & $-$0.07 & $+$0.01 & $+$0.08  &  0.11  & 0.08 \\
Fe\,{\sc ii}   &    0.00  & $+$0.08  & $-$0.03 &    0.00 & $+$0.10  &  0.13  & 0.10 \\
Na\,{\sc i}    & $+$0.03  &    0.00  & $-$0.01 &    0.00 & $+$0.13  & 0.13   & ---  \\
Mg\,{\sc i}    & $+$0.06  & $-$0.01  & $-$0.06 & $+$0.01 & $+$0.06  & 0.10   & 0.15 \\
Ca\,{\sc i}    & $+$0.05  & $-$0.01  & $-$0.05 & $+$0.01 & $+$0.07  & 0.10   & 0.08 \\
Ti\,{\sc i}    & $+$0.08  & $-$0.02  & $-$0.04 & $+$0.01 & $+$0.08  & 0.12   & 0.09 \\
Cr\,{\sc i}    & $+$0.09  & $-$0.01  & $-$0.07 & $+$0.01 & $+$0.07  & 0.13   & 0.05 \\
Co\,{\sc i}    & $+$0.10  & $-$0.05  & $-$0.10 & $+$0.00 & ---      & 0.15   & ---  \\
Ni\,{\sc i}    & $+$0.07  &    0.00  & $-$0.01 &    0.00 & $+$0.10  & 0.12   & 0.21 \\
Zn\,{\sc i}    & $+$0.06  & $+$0.04  & $-$0.03 &    0.00 & $+$0.08  & 0.11   & ---  \\
Sr\,{\sc ii}   & $+$0.06  & $+$0.01  & $-$0.23 &    0.00 & $+$0.04  & 0.25   & ---  \\
Y\,{\sc ii}    & $+$0.02  & $+$0.06  & $-$0.04 & $-$0.01 & $+$0.07  & 0.10   & 0.10 \\
Zr\,{\sc ii}   & $+$0.03  & $+$0.07  & $-$0.05 & $-$0.01 & $+$0.08  & 0.12   & 0.20 \\
Ba\,{\sc ii}   & $-$0.01  & $+$0.09  & $-$0.01 & $-$0.01 &  ---     & 0.09   &  --- \\
La\,{\sc ii}   & $-$0.03  & $+$0.06  & $-$0.06 & $-$0.01 & $+$0.12  & 0.15   & ---  \\
Ce\,{\sc ii}   & $+$0.03  & $+$0.06  & $-$0.03 & $-$0.01 & $+$0.10  & 0.12   & 0.08 \\
Pr\,{\sc ii}   & $+$0.10  & $+$0.10  & $-$0.05 & $-$0.01 & ---      & 0.15   & ---  \\
Nd\,{\sc ii}   & $+$0.04  & $+$0.06  & $-$0.02 & $-$0.01 & $+$0.09  & 0.12   & 0.11 \\
Sm\,{\sc ii}   & $+$0.03  & $+$0.07  & $-$0.01 & $-$0.01 & $+$0.14  & 0.16   & 0.10 \\
Eu\,{\sc ii}   & $+$0.01  & $+$0.09  & $-$0.20 & $+$0.01 &  ---     & 0.22   & ---  \\
Pb\,{\sc i}    &    0.10  & $-$0.03  & $+$0.05 &    0.00 &  ---     & 0.12   & ---  \\\hline
\end{tabular}
\end{table*}

\section{Discussion}

\subsection{The luminosity of CD-50$^\circ$776}

\par Once we estimated the temperature and gravity of
CD-50$^\circ$776, we are able to determine the luminosity considering
the relation
\begin{equation}
log\,(L_{\star}/L_{\odot})\,=4\log T_{\rm eff \star} - \log g_{\star} + \frac{M_{\star}}{M_{\odot}}+10.61
\end{equation}
where we considered $T$$_{\rm eff \odot}$\,=\,5\,777\,K and $\log g$$_{\odot}$\,=\,4.44.

\par Inserting the values of $T_{\rm eff}=4\,900$\,K, $\log g=2.1$,
and assuming a mass $M_{\star}=0.8M_{\odot}$ for CEMP stars (Aoki et
al. 2007), we obtain for the luminosity of CD-50$^\circ$776 a value of
log\,$(L_{\star}/L_{\odot})$\,=\,$1.95\pm0.3$.  Spectroscopic
luminosities of low-metallicity giants derived from ionization balance
may give higher values than those derived from stellar parallaxes or
evolutionary models (Mashonkina et al. 2011; Ruchti et
al. 2013). According to the recent work of Ruchti et al. (2013), the
non-local thermodynamic equilibrium (NLTE) correction to the
spectroscopic gravity is about $+1.0$\,dex, and for the temperature
the correction is around $+400$\,K.  Introducing these corrections in
equation (1), we obtain a luminosity of
log\,$(L_{\star}/L_{\odot})=1.09\pm0.3$.  In Figure~6 we show the
derived temperature and gravity of CD-50$^\circ$776 in the
log\,$T_{\rm eff}$\,-\,$\log g$ plane, together with the 12 and 14 Gyr
Yale-Yonsei isochrones for a metallicity of [Fe/H]\,=\,$-$2.2 (Kim et
al. 2002).

\par As mentioned in Section 3.1, Beers et al. (2014) also determined
the temperature, surface gravity and metallicity of CD-50$^\circ$776
using three different techniques. However, their results using
high-resolution spectroscopy provided a surface gravity $+0.9$ higher
than the value obtained by us.

\begin{figure} 
\includegraphics[width=\columnwidth]{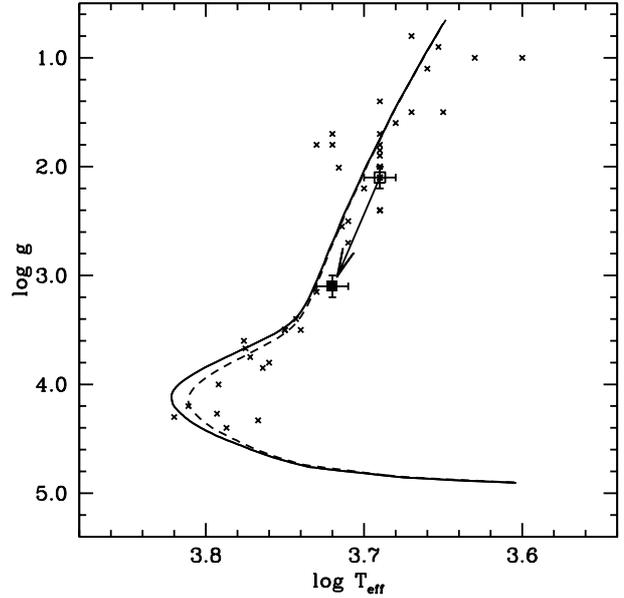}
\caption{Location of CD-50$^\circ$776 in the  $\log T_{\rm eff}$ - $\log g$ plane. 
We also show the isochrones from Kim et al. (2002)
for a metallicity of [Fe/H]\,=\,$-$2.2. The arrow corresponds to a
correction of $+400$\,K in the temperature and $+1.0$\,dex 
in the surface gravity of CD-50$^\circ$776, which is necessary due to the use of 
LTE model atmospheres (open square) instead of NLTE model atmospheres 
(filled square). Data for CEMP-s stars (black crosses) were taken from 
Aoki et al. (2002, 2007), Roederer et al. (2014),
Lucatello et al. (2003), Barklem et al. (2005), Goswami et al. (2006), 
Cohen et al. (2006), Masseron et al. (2010) and Drake \& Pereira (2011).}
\end{figure}

\subsection{CD-50$^\circ$776 as a new CEMP star}

\par In Figure~7 we reproduce Figure~4 of Aoki et al. (2007), where
the authors presented a new constraint for a star to be classified as
a CEMP star. In Figure~7a, the position of CD-50$\degr$776 is clearly
above the lower limit for a star to be considered as a CEMP
star. Figure 7b plots the [C/Fe] ratio {\sl versus.} metallicity for the
CEMP stars and again CD-50$\degr$776 occupies in this diagram the same
position as other CEMP stars. Therefore, based on these two diagrams,
we can classify CD-50$\degr$776 as a new CEMP star. In addition, the
position of CD-50$\degr$776 in figures 2 and 6 of Yoon et al. (2016)
also supports our conclusion that CD-50$\degr$776 is a CEMP star. We
will show in Section 4.4.3 that, based on the abundance analysis,
CD-50$\degr$776 is actually a new CEMP-s star.

\begin{figure} 
\includegraphics[width=\columnwidth]{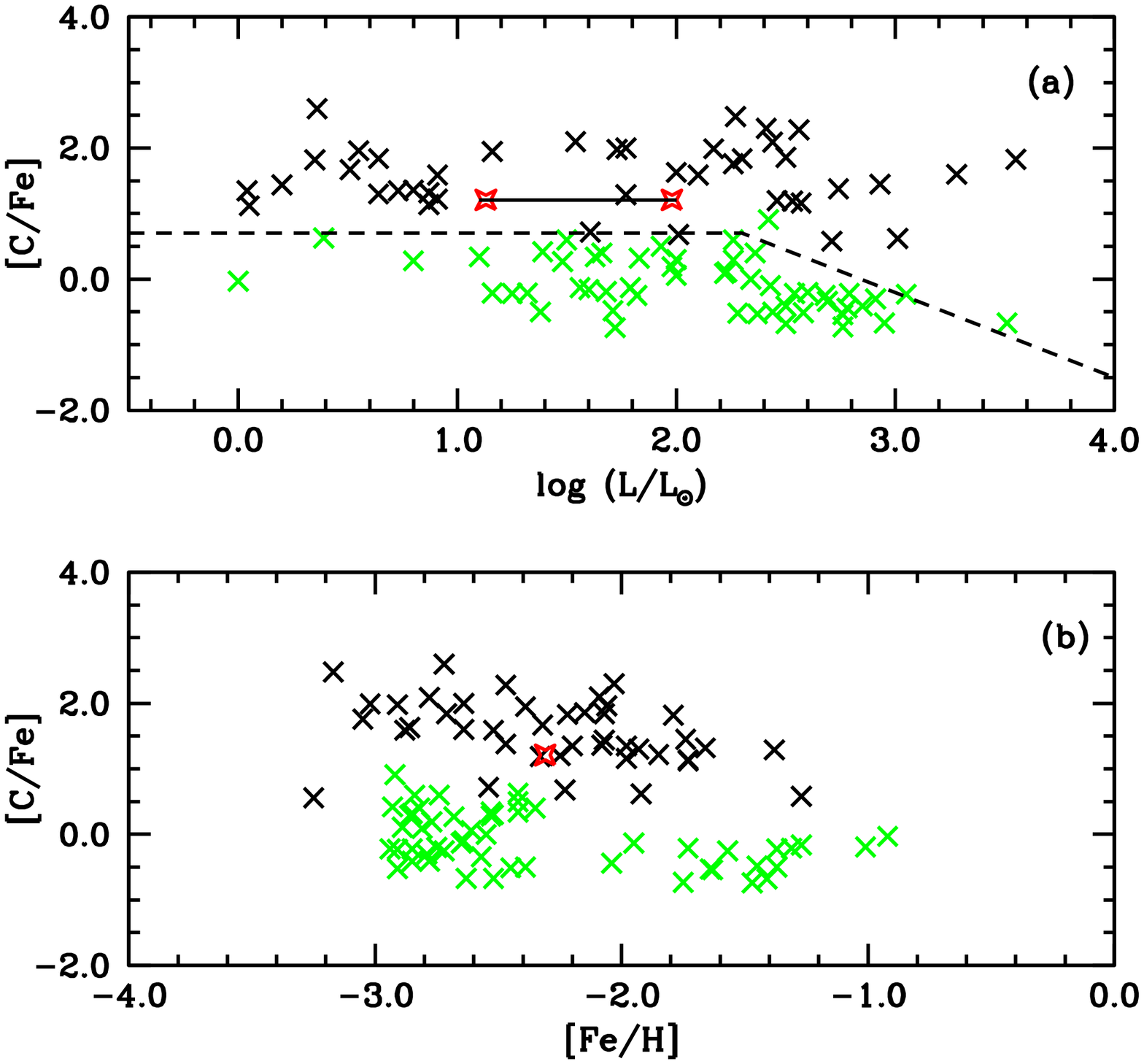}
\caption{\textbf{(a)} The [C/Fe] ratio {\sl versus.} luminosity for 
CD-50$\degr$776 (red star) in comparison with a sample of CEMP-s stars. 
The solid line connects the two possible values of luminosity of CD-50$\degr$776,
derived with and without taking into account the NLTE corrections, respectively. 
Data for CEMP-s stars are the same as in Figure~6. 
\textbf{(b)} The [C/Fe] ratio {\sl versus.} metallicity for the same stars
as in (a). The position of CD-50$\degr$776 in these two diagrams
shows that CD-50$\degr$776 is CEMP star. Green crosses represent non-carbon-rich 
metal-poor field stars. Data for these stars were taken from Gratton
et al. (2000); Cayrel et al. (2004); Honda et al. (2004) and Aoki et al. (2005).}
\end{figure}

\subsection{Radial Velocity}

\par Table 6 shows the all known measurements of the radial velocity
of CD-50$\degr$776 available in the literature and determined in this
work.  It is clear that the radial velocity of CD-50$\degr$776
presents variations due to orbital motion. Systematic radial velocity
monitoring is necessary to confirm the possible binary nature of this
new CEMP-s star.

\begin{table*} 
\caption{Radial velocity measurements for CD-50$\degr$776.}
\begin{tabular}{lcc}
\hline
Reference                &  Radial Velocity    & Modified Julian Date \\\hline 
This work (September 25, 2016) & 30.5$\pm$0.7  &  57478.17\\
This work (October 26, 1999)   & 25.2$\pm$0.8  &  57657.08 \\
Schuster et al. (2006)     & 26.0$\pm$7.0  & \\
Beers et al. (2000)        & 18$\pm$10     & \\\hline
\end{tabular}
\end{table*}

\subsection{Abundances}

\subsubsection{Nitrogen and $^{12}\mathrm{C}/\,^{13}\mathrm{C}$ isotopic ratio}

\par As shown in Table 4, the nitrogen abundance is low and the
$^{12}$C/$^{13}$C isotopic ratio is high. Combining these results with
the high carbon abundance led us to conclude that the CN cycle was not
efficient enough in the donor star of the binary system of
CD-50$\degr$776, and that significant carbon produced by the triple
alpha process was transferred in the AGB stage. The sum
(C$+$N)\,=\,7.42 illustrates the fact that carbon is the actual
responsible for the total sum of (C$+$N).

\par If we assume that the nitrogen observed in the CEMP-s stars has
the same origin as carbon, that is, it originates in the companion
star during the AGB phase (Masseron et al.  2010), then the low
nitrogen abundance observed in CD-50$\degr$776 may constrain, in
principle, the mass of the donor star.  In particular, for
CD-50$\degr$776, it is likely that the mass of the AGB star should not
have been greater than 3.0\,M$_{\odot}$. Models of AGB stars for a
metallicity of Z\,=\,0.0001 (Herwig 2004), which is the metallicity of
CD-50$\degr$776, show that the yields of carbon and nitrogen provide a
ratio [C/N] of about 2.3 for a star of 2.0\,M$_{\odot}$ at the end of
the AGB phase. Threfore, nitrogen is not enhanced in such
models. These models also predict a high $^{12}$C/$^{13}$C isotopic
ratio (figures 7 and 8 of Herwig 2004).  For a star of
3.0\,M$_{\odot}$, the ratio of [C/N] is 2.1 according to the yields
given in Herwig (2004), allowing to conclude that low metallicity
stars with masses between 2.0 and 3.0\,M$_{\odot}$ should not be
nitrogen enriched (Johnson et al. 2007).  This would explain the low
abundance of nitrogen observed in CD-50$\degr$776, and implies that
this star follows the evolution expected according to the models of
Herwig (2004). However, for other CEMP-s stars, the high nitrogen
abundance poses challenges to the evolutionary models. Masseron et
al. (2010) considered that extra mixing mechanisms should be taken
into account in order to explain the high abundance of nitrogen in
CEMP-s stars. This is because a high nitrogen abundance is predicted
by hot bottom burning, which occurs in stars with masses greater than
4.0\,M$_{\odot}$ (Sackmann \& Boothroyd 1992).  Notwithstanding, since
CEMP-s stars are members of the halo population, it is unlikely that
their companions had masses larger than 4.0\,M$_{\odot}$. This led
Masseron et al. (2010) to conclude that the nitrogen abundance in
CEMP-s stars should not be used to constrain the mass of the donor
star.

\par Figure 8 shows the [N/Fe] ratio {\sl versus.} metallicity for
CD-50$\degr$776 (red star) compared to CEMP-s giants and dwarfs
(squares), CH stars (polygons), one metal-poor barium star (HD 123396,
(1)), one CEMP-no star (CS 22877-001, (2)) and one carbon star
(HD187216, (3)).

\begin{figure} 
\includegraphics[width=\columnwidth]{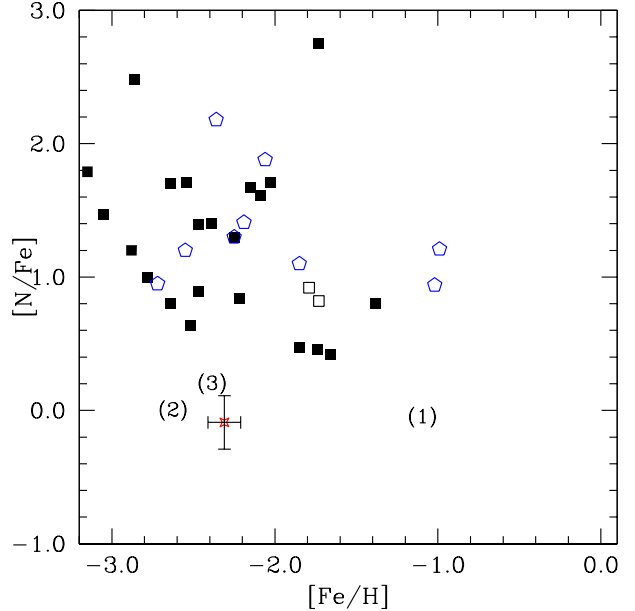}
\caption{[N/Fe] ratio {\sl versus.} metallicity for CD-50$\degr$776 (red star)
and some CEMP-s stars. Black filled squares and open squares represent 
CEMP-s giants and dwarfs, respectively. Data for CEMP-s stars are the same as
in Figure~6. Blue polygons represent CH stars with abundance data
taken from Masseron et al. (2010) and Pereira \& Drake (2009). 
Other chemically peculiar stars are also shown: HD 123396, a metal-poor
barium star ((1), Allen \& Barbuy 2006); CS 22877-001, a CEMP-no star 
((2), Masseron et al. 2010) and HD 187216, a carbon star ((3), Komiya et al. 2007).}
\end{figure}

\subsubsection{Sodium to Nickel}

\par Since CD-50$\degr$776 is a new CEMP-s star (Section 4.4.3), we
also compare its abundances to other CEMP-s stars. Figures~9 and 10
show the abundance ratios [X/Fe] {\sl versus.}  metallicity for Na,
$\alpha$-elements and iron-peak elements (Cr, Co, Ni and Zn) of
CD-50$\degr$776 compared to several previous abundance studies of
stars of the thin and thick disks and the halo populations.  CEMP-s
giant stars and dwarfs are represented by filled and open squares,
respectively. We also plot in these Figures the abundance ratios of
barium stars and of some CH stars (except carbon, cobalt, zinc and
barium), based on the recent analysis by de Castro et al. (2016).

\par Sodium abundance in CEMP-s stars exhibits the same trend as for
the other metal-poor field stars. Some CEMP-s stars display higher
[Na/Fe] ratios than the stars with similar metallicity, however this
can be caused by NLTE effects, which seem to be stronger in metal-poor
stars (see Aoki et al. 2007 for a discussion of sodium abundance in
CEMP stars). Our derived value of $+0.07$ for the [Na/Fe] ratio
indicates that NLTE effects seem to be negligible in this star.

\par In Figure~9, we verify that the abundances of $\alpha$-elements
(Mg, Ca and Ti) are those of other authors in the
field stars of the same metallicity as CD-50$\degr$776.  The
iron-group element nickel is expected to follow the iron abundance
(Figure~10), as it actually does, with a [Ni/Fe] ratio equal to
$+0.07$.  Down to [Fe/H]\,$<-2.0$, the [Ni/Fe] ratio has a scatter
around the mean [Ni/Fe]\,=\,0.0. Chromium in CD-50$\degr$776 has a
negative [Cr/Fe] ratio ($-0.17$), following the same ratio observed
in stars with equal metallicity, as well as in some CEMP-s stars.  The
other iron-peak elements, cobalt and zinc, do not deviate from the
trend observed in the field giants of the same metallicity.
Concerning the other CEMP-s stars, most of them also follow the same
trend as the metal-poor field giants for the $\alpha$-elements and the
iron-peak elements. However, some of them present high [X/Fe] ratios
for the Mg, Ca or Ti. Aoki et al. (2007) considered that some of these
high [X/Fe] ratios, like calcium for example, ``are possibly
overestimated due to contamination by molecular features'' since these
stars have low temperatures. Another possibility is that the high
[X/Fe] ratios would come from ``faint supernovae'' explosions (see
Aoki et al. 2007 and the references therein). In addition some CEMP-s
stars have abnormal low or high [X/Fe] ratios of the iron-peak
elements, specially Cr and Ni. This is probably because the abundances
of these elements in these stars were derived from one single line of
each element (Cohen et al. 2006, Aoki et al. 2007).

\begin{figure}
\includegraphics[width=\columnwidth]{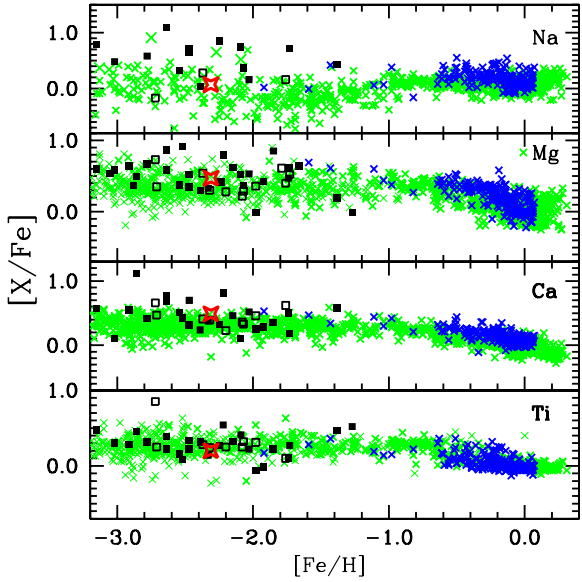}
\caption{Abundance ratios {\sl versuss.} [Fe/H] for Na, Mg, Ca and Ti.
The red star is CD-50$\degr$776. Blue crosses are barium and CH
stars from de Castro et al. (2016) and references therein; green
crosses are field giants; filled squares are CEMP-s giant stars, and open
squares are CEMP-s dwarf stars. 
Abundance ratios for field giants shown in this Figure and for the other
elements were taken from Gratton \& Sneden (1991), Gratton \& Sneden
(1994), Pilachowski et al. (1996), Burris et al. (2000), Fulbright
(2000),Mishenina \& Kovtyukh (2001), Mishenina et al. (2002), Cayrel
et al. (2004), McWilliam et al. (1995), Johnson (2002), Honda et
al. (2004), Aoki et al. (2005), Barklem et al. (2005), Mishenina et
al. (2006), Fran\c cois (2007), Mishenina et al. (2007), Luck \&
Heiter (2007), Sneden et al. (2009), Zhang et al. (2009), Alves Brito et al. (2010), For \&
Sneden (2010), Ishigaki et al. (2013) and Roederer et al. (2014). Data
for CEMP-s stars were taken from Preston \& Sneden (2001), Aoki et
al. (2002), Lucatello et al. (2003), Barklem et al. (2005), Cohen et
al. (2006), Goswami et al. (2006), Masseron et al. (2006), Aoki et
al. (2007) and Drake \& Pereira (2011).}
\end{figure}

\begin{figure}
\includegraphics[width=\columnwidth]{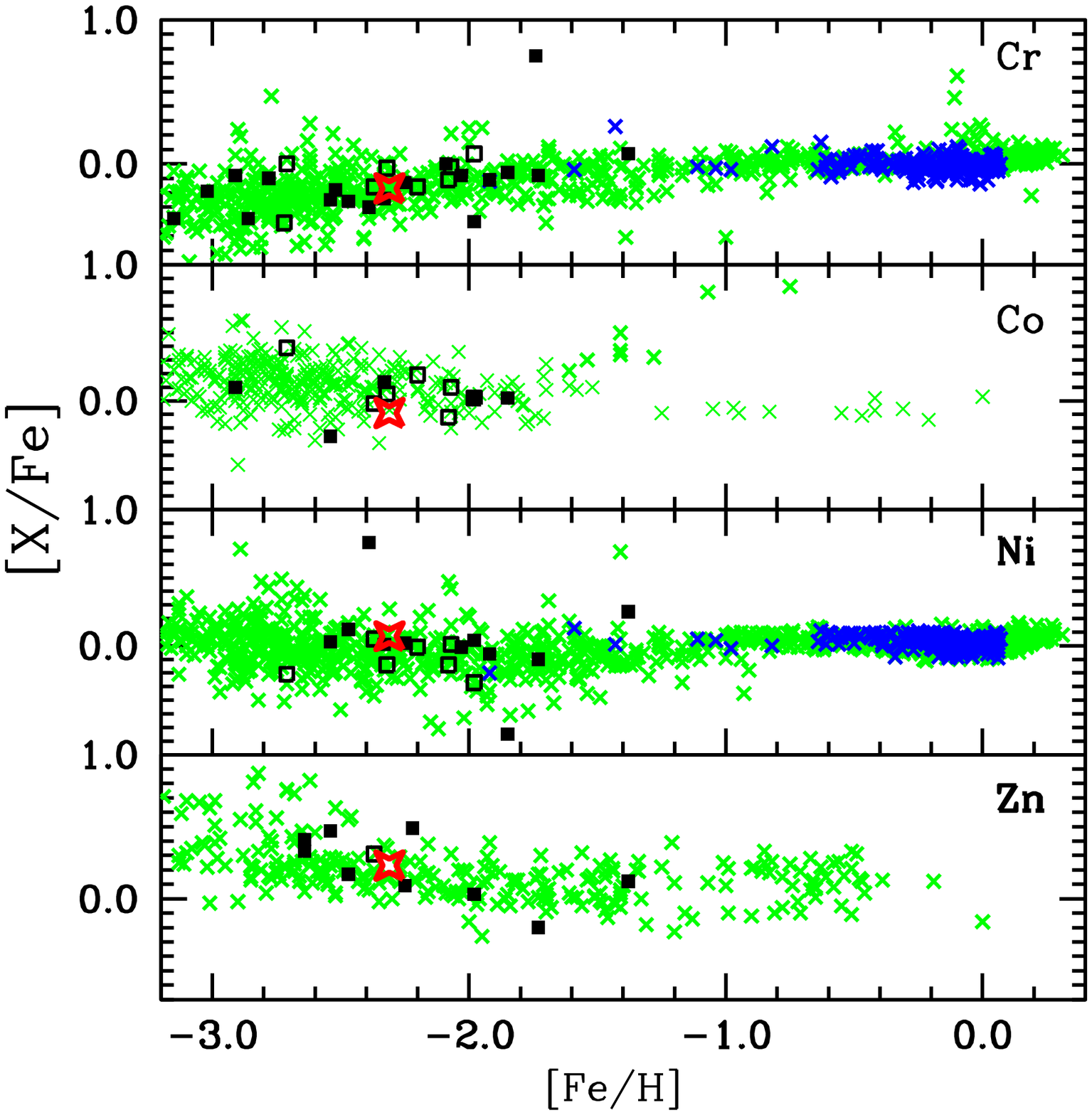}
\caption{Abundance ratios {\sl versuss.} [Fe/H] for Cr, Co, Ni and Zn.
Symbols have the same meaning as in Figure~9.}
\end{figure}

\subsubsection{The heavy-elements: CD-50$\degr$776 as a new CEMP-s star}

\par In Figures 11 and 12, we show the [X/Fe] ratios for the elements
created by the r- and s-process: Sr, Y, Zr, Ba, La, Ce, Pr, Nd, Sm and Eu, in
CD-50$\degr$776 compared to other CEMP-s stars, field giants and
barium stars, including barium stars and CH stars for several
metallicities.  Models of galactic chemical evolution do not predict
the observed overabundances of the s-process elements observed in
these plots (Travaglio et al. 1999, 2004). Since CD-50$\degr$776 also
follows the criteria given in Masseron et al. (2010) for a star to be
considered as a CEMP-s star, that is [Ba/Fe]\,$>1.0$ and
[Ba/Eu]\,$>0.0$ (our results are $+1.01$ and $+1.27$, respectively,
see Table~4), we can finally classify CD-50$\degr$776 as a new CEMP-s
star.  The {\sl mean} abundance ratio of the s-process elements
([Sr/Fe], [Y/Fe], [Zr/Fe], [Ba/Fe], [La/Fe], [Ce/Fe], [Nd/Fe] and
[Pb/Fe]) for CD-50$\degr$776 is high: $+0.77$.  If the radial velocity
variation reported in Table~6 can be attributed to orbital motion,
then the atmosphere of CD-50$\degr$776 could have been contaminated by
an extrinsic past event like in the mass-transfer hypothesis, which is
the standard scenario to explain the excess of carbon and the
overabundances of the s-process elements in these chemically peculiar
stars (Hansen et al. 2016).

\par Figures 11 and 12 also show that the abundance ratios [X/Fe] of
the light elements of the s-process are lower than those of the heavy
elements of the s-process.  This is expected based on the s-process
element production according to metallicity, since the first-peak
elements (such as Sr, Y and Zr) are bypassed in favor of the second
and third-peak elements (Busso, Gallino \& Wasserburg 1999).  Other
CEMP-s stars show the same behavior (Aoki et al. 2007).  For the
elements of the r-process, the abundance ratios of [Pr/Fe], [Sm/Fe]
and [Eu/Fe] in CD-50$\degr$776 is similar to other CEMP-s stars
previous analyzed.  In addition, the low [Eu/Fe] ratio indicates that
CD-50$\degr$776 is a CEMP-s star.

\par We note that CD-50$\degr$776 is also a ``lead
star''.  Figure~13 shows the [Pb/Ce] ratio as a function of
metallicity for CD-50$\degr$776 compared to the CH stars (blue
polygons) the CEMP-s binary stars (red circles), the barium giants
(red open squares) and the subgiant CH stars (red crosses).  The
position of CD-50$\degr$776 in this diagram, close to the CH stars and
other CEMP-s stars, indicates its lead star nature.

\begin{figure}
\includegraphics[width=\columnwidth]{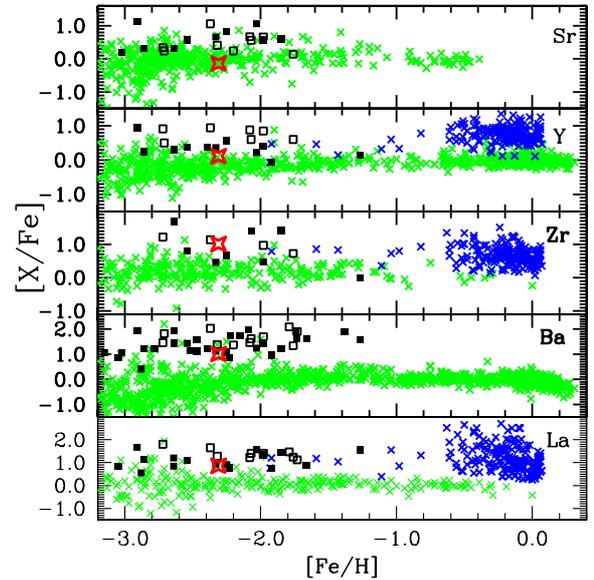}
\caption{Abundance ratios {\sl versus.} [Fe/H] for Sr, Y, Zr, Ba and La.
Symbols have the same meaning as in Figure~9.}
\end{figure}

\begin{figure}
\includegraphics[width=\columnwidth]{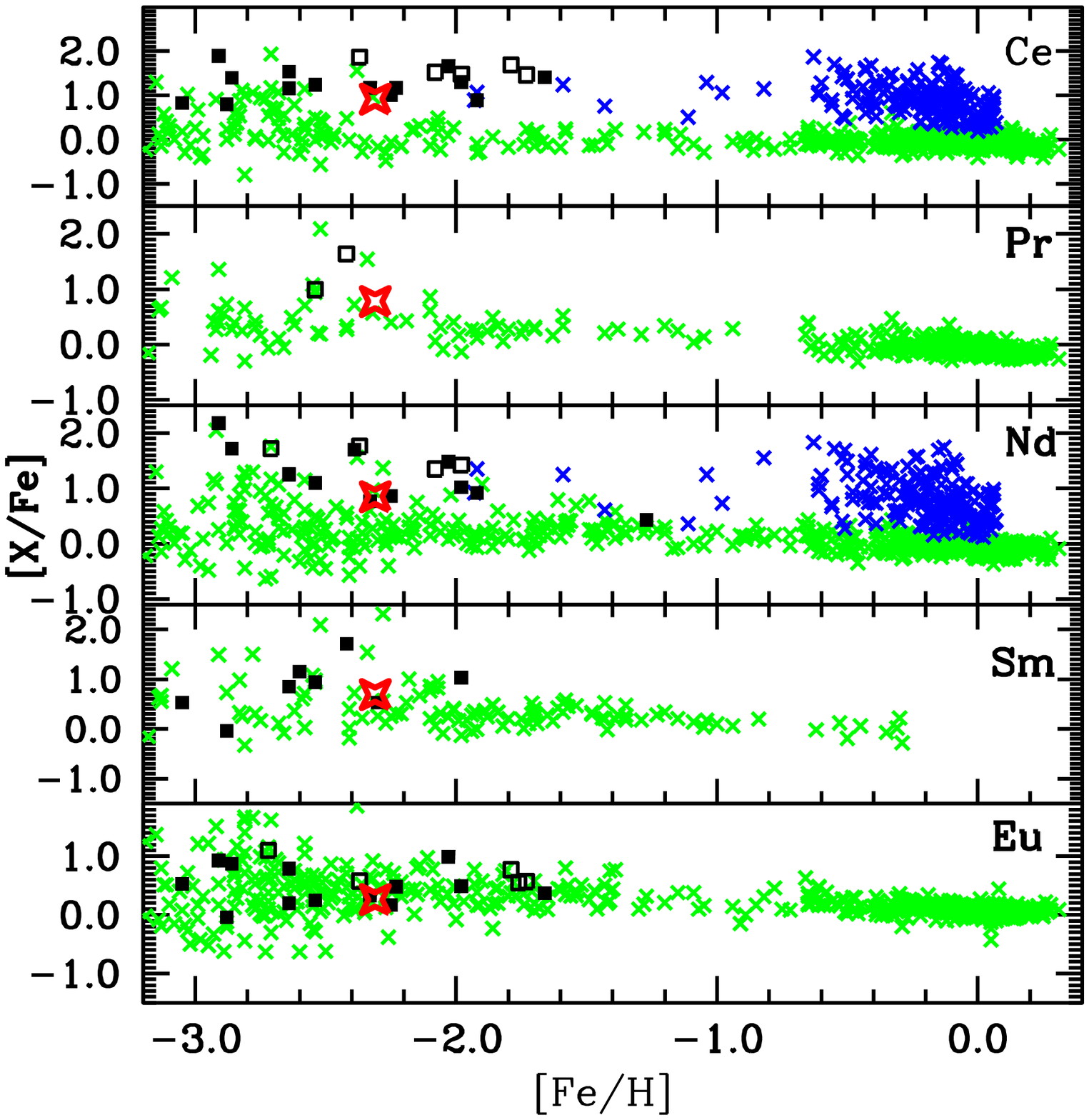}
\caption{Abundance ratios {\sl versus.} [Fe/H] for Ce, Pr, Nd, Sm and Eu.
Symbols have the same meaning as in Figure~9.}
\end{figure}

\begin{figure}
\includegraphics[width=\columnwidth]{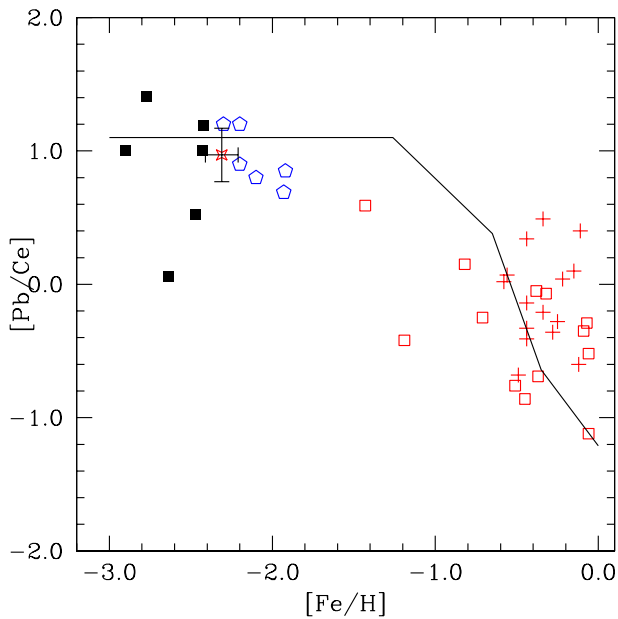}
\caption{[Pb/Ce] ratio {\sl versus.} metallicity for CD-50$\degr$776
(red star) in comparison to CEMP-s stars which are known to be binaries
(black squares), CH stars (blue polygons), barium giant stars
(red squares) and subgiant stars (red crosses). The solid line represents 
the prediction from the standard partial mixing (PM) model as given 
by Goriely \& Mowlavi (2000).}
\end{figure}

\section{Conclusions}

\par Based on high-resolution optical spectroscopic data, we present
the first detailed analysis of the chemical abundances of the CEMP
star CD-50$\degr$776, including the light elements, Na, the
$\alpha$-elements, the iron-peak elements, and the s-process elements.
We showed that CD-50$\degr$776 is characterized by an enhancement of
carbon, s-process elements, and lead.  This pattern, together with its
low metalicity ([Fe/H]\,$=-2.31$), indicates that it is CEMP-s star.
CD-50$\degr$776 is also a ``lead star'', since its lead-to-cerium
ratio $+0.97$ follows the theoretical predictions for a star of this
metallicity.

\par One way to verify that CD-50$\degr$776 is indeed a CEMP-s star is
to use the nucleosynthesis models for AGB stars calculated by Bisterzo
et al. (2010). Using the tables given in this paper, we can compare
the predicted surface abundance ratios, [X/Fe], with the observed
abundances. The nucleosynthesis models forecast the theoretical [X/Fe]
ratios for AGB stars with initial masses of 1.3M$_{\odot}$,
1.4M$_{\odot}$, 1.5M$_{\odot}$ and 2.0M$_{\odot}$, varying the number
of thermal pulses and the quantity of $^{13}$C pocket for a
metallicity [Fe/H]\,=\,$-$ 2.6, close to the metallicity of
CD-50$\degr$776.

\par Figure 14 illustrates this comparison, and shows that the best
nucleosynthesis model that fits the observations is that of a star
with an initial AGB mass of 1.3M$_{\odot}$ for the ST/2
case. Inspecting another fits for the CEMP-s stars investigated in
Bistezro et al. (2012), we verify that the abundance pattern of
CD-50$\degr$776 is similar to the pattern of the CEMP-s stars CS
22964-161, CS 22880-074, CS 22942-019, CS 30301-015, HD 196944, and BS
17436-058, where the abundance of lead was also determined.  These
stars were classified by Bistezro et al. (2012) as CEMP-sI, which
means that the ratio [hs/Fe] (defined by Bistezro et al. (2012) as the
mean [X/Fe] ratio given by ([La/Fe]$+$[Nd/Fe]$+$[Sm/Fe])/3.) is less
than 1.5. In fact, CD-50$\degr$776 has [hs/Fe]\,=\,0.8.

\par However, CD-50$\degr$776 presents another chemical peculiarity
rarely observed in the CEMP-s stars, that is, a low abundance of
nitrogen. As far as we know, this peculiarity has also been observed
in the extragalactic CEMP-s star Scl-1013644 (Salgado et al. 2016). It
is worth noting that the nucleosynthesis models of Bisterzo et
al. (2010) predict a high abundance of carbon and nitrogen for
CD-50$\degr$776, which is not supported by our observations.

\par As mentioned in Bisterzo et al. (2011), the ratios of [C/Fe],
     [N/Fe] and the $^{12}$C/$^{13}$C isotopic ratios are
     overestimated in AGB models where the occurrence of mixing
     produced by the 'Cool Bottom Processing' (CBP) has been accounted
     to explain the abundances of carbon and nitrogen in CEMP-s stars.
     However, the efficiency of this process is difficult to estimate
     due to the influence of other physical phenomena such as
     rotation, thermohaline mixing, and magnetic fields.

\par On the other hand, as discussed in Section 4.1.1, the low
abundance of nitrogen could be explained assuming an initial mass of
2.0M$_{\odot}$ of the donor star without the occurrence of CBP. Thus,
it seems that the mixing process and their efficiency in both the AGB
star and the star that received the ejected material should be better
modeled to fit the observations.  Finally, we recall that further
spectroscopic observations will be important to obtain radial velocity
measurements and to investigate the binary nature of this star.

\begin{figure}
\includegraphics[width=\columnwidth]{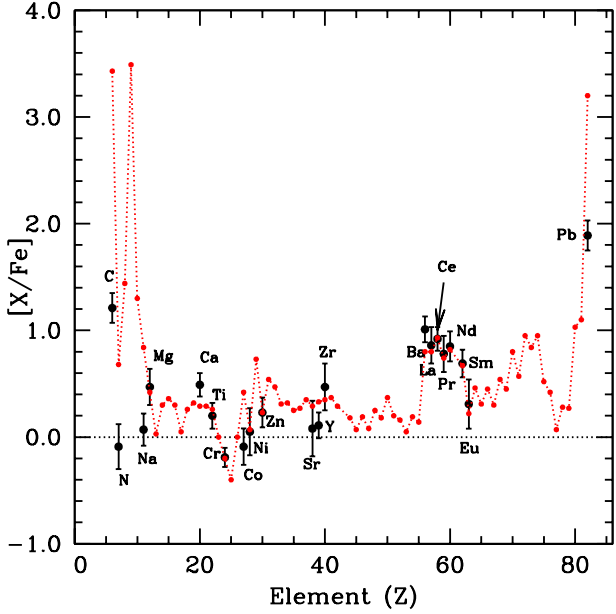}
\caption{Observed (black dots) and predicted (red) [X/Fe] ratios for 
CD-50$\degr$776. The predicted [X/Fe] ratios were obtained using the models 
available in Bistezo et al. (2010). The best model was obtained for a initial 
AGB star for 1.3M$_{\odot}$, case ST/2 after five thermal pulses and no dilution.}
\end{figure}

\section{Acknowledgements}

\par N.A.D. acknowledges FAPERJ, Rio de Janeiro, Brazil, for Visiting
Researcher grant E-26/200.128/2015 and the Saint Petersburg State
University for research grant 6.38.335.2015. We also thank the referee,
Chris Sneden, for the valuable remarks that improved the paper.

{}

\bsp
\label{lastpage}

\begin{thebibliography}{}

\bibitem{}
Allen, D.M. \& Barbuy, B. 2006, A\&A, 454, 895        

\bibitem{}
Alves-Brito, A., Mel\'endez, J., Asplund, M., Ram{\'i}rez, I., Yong, D.C., 2010, 
A\&A, 513, 35     

\bibitem{}
Aoki, W., Norris, J.E., Ryan, S.G., Beers, T.C. \& Ando, H., 2002, \apj, 567, 1166   

\bibitem{}
Aoki, W., Honda, S. Beers, T.C., Kajino, T. Ando, H., et al. 2005, \apj, 632, 611     

\bibitem{}
Aoki, W., Beers, T.C., Christlieb, N., Norris, J.E., Ryan, S.G. \& Tsangarides, S., 2007,
\apj, 655, 492      

\bibitem{}
Barbuy, B., Cayrel, R., Spite, M., Beers, T.C. Spite, F. et al, 1997, A\&A, 317, L63    

\bibitem{}
Barklem, P.S., Christlieb, N., Beers, T.C., Hill, V., Bessell, M.S. et al. 2005, A\&A,    
439, 129   

\bibitem{}
Beers, T. C., Preston, G. W. \& Shectman, S.A., 1985, \aj, 90, 2089   

\bibitem{}
Beers, T.C., Chiba, M., Yoshii, Y., Platais, I., Hanson, R.B., 2000, \aj, 119, 2866   

\bibitem{}
Beers, T.C., Preston, G.W. \& Shectman, S.A., 1992, \aj, 103, 1987   

\bibitem{}
Beers, T.C. \& Christlieb, N., 2005, \araa, 43, 531   

\bibitem{}
Beers, T.C., Norris, J.E., Placco, V.M., Lee, Y.S., Rossi, S., Carollo, D. \& 
Masseron, T., 2014, \apj, 794, 58     

\bibitem{}
Beers, T.C., Placco, V.M., Carollo, D., Rossi, S., Lee, Y.S. et al., 2017, \apj, 835, 81   

\bibitem{}
Biemont, E. \& Godefroid, M., 1980, A\&A, 84, 361     

\bibitem{}
Bidelman, W. P., 1981, \aj, 86, 553   

\bibitem{}
Bidelman, W.P. \& MacConnell, D.J., 1973, \aj, 78, 687    

\bibitem{}
Bisterzo, S., Gallino, R., Straniero, O., Cristallo, S. \& K\"appeler, F., 2010, \mnras, 404, 1529   

\bibitem{}
Bisterzo, S., Gallino, R., Straniero, O., Cristallo, S. \& K\"appeler, F., 2011, MNRAS, 418, 248   

\bibitem{}
Bisterzo, S., Gallino, R., Straniero, O., Cristallo, S. \& K\"appeler, F., 2012, MNRAS, 422, 849   

\bibitem{}
Bond, H., 1970, ApJS, 22, 117   

\bibitem{}
Bond, H., 1980, ApJS, 44, 517   

\bibitem{}
Bonifacio, P., Molaro, P., Beers, T.C. \& Vladilo, G., 1998, A\&A, 332, 672   

\bibitem{}
Burris, D.L., Pilachowski, C.A., Armandroff, T.E., Sneden, C. \& Cowan, J.J.
2000, \apj, 544, 302     

\bibitem{}
Busso, M., Gallino, R. \& Wasserburg, G.J., 1999, \araa, 37, 239    

\bibitem{}
Cayrel, R., Depagne, E., Spite, M., Hill, V.; Spite, F., 2004, A\&A, 416, 1117     

\bibitem{}
Chen, Y.Q., Zhao, G., Nissen, P.E., Bai, G.S. \& Qiu, H.M., 2003, ApJ, 591, 925     

\bibitem{}
Christlieb, N., Green, P.J., Wisotzki, L. \& Reimers, D., 2001, A\&A,, 375, 266   

\bibitem{}
Cohen, J.G., McWilliam, A., Shectman, S., Thompson, I., Christlieb, N. et al. 2006,
\apj, 132, 137     

\bibitem{}
de Castro, D.B., Pereira, C.B., Roig, F., Jilinski, E., Drake, N.A. et al. 2016, \mnras,
459, 4299    

\bibitem{}
Depagne, E., Hill, V., Spite, M., Spite, F., Plez, B., et al. 2002, A\&A, 390, 187        

\bibitem{}
Den Hartog, E.A., Lawler, J.E., Sneden, C. \& Cowan, J.J., 2003, ApJS, 148, 543    

\bibitem{}
Drake, J.J. \& Smith, G., 1991,  MNRAS, 250, 89           

\bibitem{}
Drake, N.A. \& Pereira, C.B., 2011, A\&A, 531, 133     

\bibitem{}
Edvardsson, B., Andersen, J., Gustafsson, B., Lambert, D.L., Nissen, P. E.,
et al, 1993, A\&A, 275, 101     

\bibitem{}
For, Bi-Qing \& Sneden, C., 2010, \aj, 140, 1694     

\bibitem{}
Fran\c cois, P., Depagne, E., Hill, V., Spite, M., Spite, F., et al, 2017, A\&A,, 476,
935     

\bibitem{}
Frebel, A., \& Norris, J. E. 2013, in Planets, Stars, and Stellar Systems, Vol. 5,
ed. T. Oswalt \& G. Gilmore (Dordrecht: Springer), p.5    

\bibitem{}
Fulbright, J.P., 2000, \aj, 120, 1841          

\bibitem{}
Goriely, S. \& Mowlavi, N., 2000, A\&A, 362, 599        

\bibitem{}
Goswami, A., A  i, W., Beers, T.C., Christlieb, N., Norris, J.E. et al. 2006, 
\mnras, 372, 343     

\bibitem{}  
Gratton, R.G. \& Sneden, C., 1988, A\&A, 204, 193        

\bibitem{}  
Gratton, R.G. \& Sneden, C., 1991, A\&A,, 241, 501   

\bibitem{}
Gratton, R.G. \& Sneden, C., 1994, A\&A< 287, 927    

\bibitem{}
Gratton, R.G., Sneden, C., Carretta, E. \& Bragaglia, A., 2000, A\&A,, 354, 169   

\bibitem{}
Grevesse, N. \& Sauval, A.J., 1998, Spa. Sci. Rev., 85, 161        

\bibitem{}
Hannaford, P., Lowe, R.M., Grevesse, N., Biemont, E. \& Whaling, W., 1982, ApJ, 261, 736    

\bibitem{}
Hansen, C.J., Nordstr\"om, B., Hansen, T.T., Kennedy, C.R., Placco, V.M., 2016, A\&A, 
588, 3     
   
\bibitem{}
Herwig, F., 2004, ApJS, 155, 651   

\bibitem{}
Hill, V., Barbuy, B., Spite, M., Spite, F., Cayrel, R. et al, 2000, A\&A, 353, 557   

\bibitem{}
Hill, V., Plez, B., Cayrel, R., Beers, T.C., Nordstr\"om, B., 2002, A\&A, 387, 560   

\bibitem{}
Honda, S., Aokii, W., Kajino, T., Ando, H., Beers, T.C. et al. , 2004, \apj, 607, 474     

\bibitem{}
Ishigaki, M.N., Aokii, W. \& Chiba, M., 2013, \apj, 771, 671     

\bibitem{}
Johnson, J.A., 2002, ApJS, 139, 219     

\bibitem{}
Johnson, J.A., Herwig, F., Beers, T.C. \& Christlieb, N., 2007, \apj, 658, 1203   

\bibitem{}
Jonsell, K., Barklem, P.S., Gustafsson, B., Christlieb, N., Hill, V., 2006, A\&A, 451, 651   

\bibitem{} 
Kaufer, A., Stahl, O. Tubbesing, S.,  et al. 1999, The Messenger, 95, 8.            

\bibitem{}
Kim, Yong-Cheol, Demarque, P., Yi, S.K. \& Alexander, D.R., 2002, ApJS, 143, 499     

\bibitem{}
Komiya, Y., Suda, T., Minaguchi, H., Shigeyama, T., Aokii, W., et al. 2007, \apj, 658, 267    

\bibitem{}
Kurucz, R.L. 1993, CD-ROM 13, Atlas9 Stellar Atmosphere Programs and 2 km/s
Grid (Cambridge: Smithsonian Astrophys. Obs)        

\bibitem{}
Lambert, D.L., Smith, V.V. \& Heath, J., 1993, PASP, 105, 568    

\bibitem{} 
Lambert, D.L., Heath, J.E., Lemke, M. \& Drake, J., 1996, ApJS, 103, 183     

\bibitem{}
Lawler, J.E., Bonvallet, G. \& Sneden, C., 2001, ApJ, 556, 452     

\bibitem{}
Lawler, J.E., Den Hartog, E.A., Sneden, C. \& Cowan, J.J., 2006, ApJS, 162, 227.   

\bibitem{}
Lawler, J.E., Sneden, C., Cowan, J.J., Ivans, I.I. \& Den Hartog, E.A., 2009,   
ApJS, 182, 51   

\bibitem{}
Ljung, G., Nilsson, H., Asplund, M. \& Johansson, S., 2006, A\&A, 456, 1181   

\bibitem{}
Lucatello, S., Gratton, R., Cohen, J.G., Beers, T.C., Christlieb, N. et al., 2003, AJ, 
125, 875        

\bibitem{}
Lucatello, S., Beers, T.C., Christlieb, N., Barklem, P.S.; Rossi, S., 2006, \apj, 652, L37   

\bibitem{}
Luck, R.E. \& Heiter, U., 2007, AJ, 133, 2464        

\bibitem{}
Martin, W. C., Fuhr, J. R., Kelleher, D. E., et al. 2002, NIST Atomic Spectra
Database (Version 2.0; Gaithersburg, MD: NIST)     

\bibitem{}
Mashonkina, L., Gehren, T., Shi, J.-R.,  Korn, A.J. \& Grupp, F., 2011, A\&A, 528, 87   

\bibitem{}
Masseron, T., van Eck, S., Famaey, B., Goriely, S., Plez, B. et al. 2006, A\&A, 455, 1059     

\bibitem{}
Masseron, T., Johnson, J.A., Plez, B., van Eck, S., Primas, F. et al. 2010, A\&A
509, 93         

\bibitem{}
McWilliam, A., Preston, G.W., Sneden, C. \& Shectman, S., 1995, \aj, 109, 2736     

\bibitem{}
McWilliam, A. 1998, \aj 115, 1640     

\bibitem{}
Mishenina, T.V. \& Kovtyukh, V.V., 2001, A\&A, 370, 951     

\bibitem{}
Mishenina, T.V., Kovtyukh, V.V., Soubiran, C., Travaglio, C. \& Busso, M., 2002,
A\&A, 396, 189     

\bibitem{}
Mishenina, T.V., Bienaym\'e, O., Gorbaneva, T.I., Charbonnel, C., et al. 2006, A\&A, 456, 
1109    

\bibitem{}
Mishenina, T.V., Gorbaneva, T.I., Bienaym\'e, O., Soubiran, C., Kovtyukh, V.V.
et al. 2007, ARep, 51, 382     

\bibitem{}
Mucciarelli, A., Caffau, E., Freytag, B., Ludwig, H.-G. \& Bonifacio, P., 2008, A\&A, 484, 841   

\bibitem{}
Norris, J., Bessell, M.S. \& Pickles, A.J., 1985, ApJS, 58, 463     

\bibitem{}
Norris, J.E., Ryan, S. G. \& Beers, T.C., 1996, ApJS, 107,391    

\bibitem{}
Norris, J.E., Ryan, S.G. \& Beers, T.C., 1997, \apj, 488, 250   

\bibitem{}
Norris, J.E., Ryan, S.G. \& Beers, T.C., 1997, \apj, 489, L169   

\bibitem{}
Norris, J.E., Ryan, S.G. \& Beers, T.C., 2001, ApJ, 561, 1034    

\bibitem{}
Pereira, C.B. \& Drake, N.A., 2009, A\&A, 496, 791      

\bibitem{}
Pereira, C.B., Jilinski, E., Drake, N.A., de Castro, D.B., Ortega, V.G. et al. 2012,
A\&A, 543, 59     

\bibitem{}
Pereira, C. B., Jilinski, E.G., Drake, N.A., Ortega, V.G. \& Roig, F., 2013, A\&A, 559, 12     

\bibitem{}
Pilachowski, C.A., Sneden, C. \& Kraft, R.P., 1996, \aj, 111, 1689     

\bibitem{}
Preston, G.W. \& Sneden, C., 2001, ApJ, 122, 1545        

\bibitem{}
Reddy, B. E., Bakker, E. J. \& Hrivnak, B.J. 1999, ApJ, 524, 831    

\bibitem{}
Reddy, B.E., Tomkin, J., Lambert, D.L. \& Allende Prieto, C., 2003, MNRAS, 340, 304      

\bibitem{}
Roederer, I.U., Preston, G.W., Thompson, I.B., Shectman, S.A. \& Sneden, C., 2014,
\aj, 147, 136    

\bibitem{}
Rossi, S., Beers, T.C., Sneden, C., Sevastyanenko, T., Rhee, J. \& Marsteller, B., 2005, \aj, 130, 2804   

\bibitem{}
Ruchti, G.R., Bergemann, M., Serenelli, A.,  Casagrande, L. \& Lind, K., 2013, \mnras, 429, 126   

\bibitem{}
Ryan, S.G. \& Deliyannis, C.P., 1998, \apj, 500, 398    

\bibitem{}
Sackmann, I.-J. \& Boothroyd, A.I., 1992, \apj, 392, L71   

\bibitem{}
Salgado, C.; Da Costa, G.S., Yong, D. \& Norris, J.E., 2016, \mnras, 463, 598   

\bibitem{}
Schuster, W.J., Moitinho, A., M\'arquez, A., Parrao, L., Covarrubias, E., 2006, A\&A, 445, 939   

\bibitem{}
Sivarani, T., Bonifacio, P., Molaro, P., Cayrel, R., Spite, M., 2004, A\&A, 413, 1073   

\bibitem{}
Smith, V.V., Cunha, K., Jorissen, A. \& Boffin, H.M.J. 1996, A\&A, 315, 179      

\bibitem{} 
Sneden, C., 1973, Ph.D. Thesis, Univ. of Texas        

\bibitem{}
Sneden, C., Preston, G.W., McWilliam, A. \& Searle, L., 1994, \apj, 431, L27   

\bibitem{}
Sneden, C., McWilliam, A., Preston, G.W., Cowan, J.J., Burris, D.L. \& Armosky, B.J., 1996, \apj, 467, 819   

\bibitem{}
Sneden, C., Cowan, J.J., Lawler, J.E., Ivans, I.I., Burles, S., 2003a, \apj, 591, 936   

\bibitem{}
Sneden, C., Preston, G.W. \& Cowan, J.J., 2003b, \apj, 592, 504   

\bibitem{}
Sneden, C., Lawler, J.E., Cowan, J.J., Ivans, I.I. \& Den Hartog, E.A., 2009, ApJS, 182, 80   

\bibitem{}
Sneden, C., Lucatello, S., Ram, R.S., Brooke, J.S.A. \& Bernath, P., 2014, ApJS, 214, 26   

\bibitem{}
Sobeck, J.S., Kraft, R.P., Sneden, C,, Preston, G.W., Cowan, J.J. et al., 2011, \aj, 141, 175    

\bibitem{}
Travaglio, C., Galli, D., Gallino, R., Busso, M. \& Ferrini, F., 1999, ApJ, 510, 325     

\bibitem{}
Travaglio, C., Gallino, R., Arnone, E., Cowan, J.,  Jordan, F., 2004, ApJ, 601, 964        

\bibitem{}
Van Eck, S. \& Jorissen, A., 1999, A\&A, 345, 127      

\bibitem{}
van Eck, S., Goriely, S., Jorissen, A. \& Plez, B., 2003, A\&A, 404, 291      

\bibitem{}
Wood, M.P., Lawler, J.E., Sneden, C. \& Cowan, J.J., 2014, ApJS, 211, 20   

\bibitem{}
Yoon, J., Beers, T.C., Placco, V. M., Rasmussen, K.C., Carollo, D. 2016, \apj, 833, 20    

\bibitem{}
Zhang, L., Ishigaki, M., Aoki, W., Zhao, G. \& ; Chiba, M., 2009, \apj, 706, 1095     

\end{thebibliography}
\end{document}